\documentclass[preprint,12pt,usenames,dvipsnames,floats,showpacs]{elsarticle}

\usepackage{amssymb}
\usepackage{amsmath}
\usepackage{graphicx}
\usepackage[colorlinks=true, allcolors=blue]{hyperref}
\usepackage{setspace}
\usepackage{xcolor}
\usepackage{latexsym}
\usepackage{dcolumn}
\usepackage{amsmath}
\usepackage{epsf}
\usepackage{float}
\usepackage{enumerate}
\usepackage{enumitem}
\usepackage{graphicx}
\usepackage{dcolumn}
\usepackage{bm}
\usepackage{subcaption}
\usepackage{setspace}
\definecolor{mynavy}{RGB}{18,18,155}
\usepackage{soul}
\setstcolor{red}

\usepackage{amssymb}

\journal{Physics of the Dark Universe}

\begin{document}

\begin{frontmatter}



\title{A bound on the cosmic opacity of unparticles from the CMB temperature}


\author[inst1]{Maurice H.P.M. van Putten}

\affiliation[inst1]{organization={Department of physics and astronomy},
            addressline={Sejong University, 98 Gunja-Dong, Gwangjin-gu}, 
            city={Seoul},
            postcode={143-747}, 
            country={South Korea}}
            
\author[inst1,inst2]{Maryam Aghaei Abchouyeh}
\affiliation[inst2]{organization={Department of Physics, School of Natural Science},
            addressline={Ulsan National University of Science and Technology (UNIST)}, 
            city={Ulsan},
            postcode={44919}, 
            country={South Korea}}

\begin{abstract}
Unparticle cosmology gives an unconventional outlook on the dark sector of cosmology, increasingly challenging $\Lambda\mbox{CDM}$ by $H_0$-tension. This model derives from a finite temperature broken conformal symmetry of radiation, described by a non-radiative correction with unknown sign in energy density. This symmetry breaking has a sign ambiguity, in which corrections about the IR fixed point are normal or tachyonic. While the first has been ruled out in a recent study, the latter possesses a late-time temperature $T_c\simeq 4\,T_{CMB}$ associated with $\Omega_{\cal U}\simeq 1$ ($\Omega_{\cal U} \simeq 10^4 \Omega_{CMB}$), where $T_{CMB}$ denotes the temperature of the Cosmic Microwave Background (CMB). 
{Thus exposed to the enormous heat bath of unparticles, $T_{CMB}$ is constrained by the astronomical age of the Universe $T_{u,0}$.
The relative temperature shift $\Delta T/T$ in any heat exchange to the CMB  is hereby limited to the uncertainty of a few percent in $T_{u,0}$. 
This bounds the effective cross-section of unparticles interactions with CMB photons to $\sigma _{\gamma \mathcal{U}} \lesssim 10^{-40}\ \mbox{m}^2=10^{-3}\ \mbox{nb}$.} 
{Assuming a standard cross-section across the broadband mass and energy spectrum of unparticles, a more stringent bound $\sigma _{\gamma \mathcal{U}} \lesssim 10^{-3}\,\mbox{pb}$ derives from COBE-FIRAS constraints on spectral distortions.}
{The first is midway of photon-neutrino $\sigma_{\gamma\nu}$ and photon-photon cross sections $\sigma_{\gamma\gamma}$, the second is on par with $\sigma_{\gamma\nu}$. These constraints put late-time unparticle cosmology, if present, at the edge of the standard model.}
\end{abstract}


\end{frontmatter}


\section{Introduction}\label{sec:intro}

The observed accelerated Hubble expansion of the Universe is increasingly recognized as a potential challenge to $\Lambda\mbox{CDM}$ alongside other challenges for the model e.g. missing satellites, Baryon Acoustic Oscillations, the $\sigma _8$ growth tension and the like \cite{2105.05208}. While the model reliably describes the early times cosmology, where the cosmological constant (or dark energy) content is negligible, it appears unsatisfactory in explaining the late time observational results. It has stimulated researchers to introduce extended gravitational theories or new particles to account for tension in the Hubble parameter $H_0$ in $\Lambda\mbox{CDM}$ and the Local Distance Ladder (LDL) \cite{1907.10625,2001.03624,2112.04510} and the reliability of these theoretical models can be assessed by suitably chosen observational constraints. Unparticle cosmology is one attempt that holds some promise to describe the low energy Universe by introducing an additional energy density outside the realm of the standard model (SM) of particle physics and might relax at least some of this $H_0$-tension, \cite{MM,0703260,0704.2457,1301.0623,1512.05356,1709.05944}.

Unparticles represent the low energy phase of Banks-Zaks (BZ) field, first introduced by Georgi in 2007 \cite{0703260}. BZ fields possess a non-trivial IR fixed point in their interaction with SM particles at high energies above $\mathcal{M}_{\mathcal{U}}$. Below the energy scale $\mathcal{M}_{\mathcal{U}}$, their interaction with SM particles is suppressed by a power of $\mathcal{M}_{\mathcal{U}}$. Further below the energy scale $\Lambda_{\mathcal{U}}\ll \mathcal{M}_{\mathcal{U}}$, a new phase emerges as unparticles with the same properties as BZ fields along with scale invariance - unparticle stuff with no definite mass.

By definition \cite{0703260}, however, unparticles retain small interactions with SM particles, i.e. they are not completely decoupled from them. Nevertheless, in the literature it is common to assume that these interactions can be safely ignored without quantitative assessment \cite{0809.0977,artimovski,2010.02998}. Rather than an ad hoc assumption, it should be confronted with suitable observational or experimental constraints. Such may be found in late-time cosmology, highlighted in what follows.

We exploit the consistency between the age of the Universe inferred from the CMB in $\Lambda\mbox{CDM}$ and the astronomical age defined by the oldest objects in glublar clusters. 
In $\Lambda\mbox{CDM}$, this consistency rests on an effective decoupling of the CMB and the dark sector that must be preserved in the face of unparticles. 
Our analysis of unparticle cosmology hereby puts an approximate bound on the opacity and consequently the interactions of unparticles with CMB.

This paper is organized as follows: \S\ref{sec1n} introduces aforementioned {astronomical age constraint, exploited in this work to constrain late-time unparticle cosmology.}
In \S\ref{sec2n} we consider heat transfer between the CMB photons and other species. The main features and properties of unparticle cosmology is presented in \S\ref{Sec3n}. 
In \S\ref{Sec4n} we formulate the problem of suppressing heat transfer from unparticles to the CMB in terms of an {effective cross section with minimal assumptions.}
In \S\ref{Sec5n}, the effective cross section of unparticles with CMB photons is considered in two different cases, defined by bounds in the relative temperature shift in the CMB set by uncertainty in above-mentioned astronomical age of the Universe. 
{Potentially tighter constraints derived from COBE-FIRAS bounds on spectral distortions, upon assuming a standard cross section
that is effectively constant over the mass and energy spectrum of unparticles, is included in a summary of our findings in \S\ref{Sec6n}.}

\section{Two independent age constraints of the Universe \label{sec1n}}

Measuring the age of the Universe has been an interesting question for a long time. Two measurements stand out, the age derived from the CMB in $\Lambda\mbox{CDM}$ and the age of the oldest stars in globular clusters.

Planck gives us a measurement for the age of the Universe according to its $\Lambda\mbox{CDM}$-analysis of the CMB \cite{1807.06209},
\begin{equation}
    \label{ageH}
    T_{u,0}=\dfrac{1}{H_0}\int _{0}^{\infty} \dfrac{dz}{h(z)(1+z)} = \dfrac{1}{H_0}(1-\epsilon) \simeq 13.8 \pm 0.02\ \mbox{Gyr}.
\end{equation}
Here, $H(z)=H_0 h(z)$ with $h(z)=\sqrt{1-\Omega_{m,0}+\Omega_{m,0}(1+z)^3}$ parameterized by the dimensionless matter content $\Omega_{m,0}$. 
The parameter $\epsilon$ in Eq.\eqref{ageH} is $\lesssim5\%$, representing a first order correction to  $T_{u,0}\approx1/H_0$. 

Planck is fairly unambiguous on Eq.\eqref{ageH}, despite the fact that it lists various values for these late-time parameters $(H_0,\ \Omega_{m,0})$ at $z=0$. 
Uncertainty for $T_{u,0}$ is about $0.1\%$, yet uncertainty in $H_0$ is about $1\%$, while uncertainty in $\Omega_{m,0}$ is a few percent. 
This can be attributed to some degeneracy in $(H_0,\Omega_{m,0})$ in the integrand of Eq.\eqref{ageH}, given $H_0h(z)\propto H_0\Omega_{m,0}^{1/2}$ for $z\gg1$. 
The uncertainty in $\Omega_{m,0}$ hereby tends to be twice that in $H_0$, since $\delta \Omega_{m,0}/\Omega_{m,0}+ 2\delta H_0/H_0\cong 0$ ($\delta T_u/T_u \cong 0$). 
Regardless, our focus here is strictly on $T_{u,0}$, rather than $H_0$ (and $\Omega_{m,0}$) independently.

From another point of view, Planck results show that the CMB temperature $T_{CMB}$ at present satisfies $T_{CMB,0}=2.73\ \mbox{K}$ \cite{1807.06209}. It derives from the expansion of the Universe since decoupling from the surface of last scattering at $3000\ \mbox{K}$ ($z\approx1100$), corresponding to $378,000$ years after the Big Bang \cite{Kolb}. In the adiabatic limit a largely matter dominated universe gives
\begin{equation}
    \label{decoupling}
    \dfrac{T_{CMB,0}}{T_{CMB}(1100)}= \dfrac{a(1100)}{a_0}\simeq \left (\dfrac{T_u(1100)}{T_{u,0}}\right )^{2/3},
\end{equation}
where $a=a(z)$ denotes the Friedmann scale factor, $T_u=T_u(z)$, $T_{CMB}=T_{CMB}(z)$ and the subscript ``$0$'' refers to $z=0$ as before.
Accordingly, the CMB serves as an accurate tracer of late-time $\Lambda\mbox{CDM}$ evolution, indicating
\begin{equation}
    \label{ageT}
    T_{u,0}\simeq 13.7\ \mbox{Gyr}. 
\end{equation}

Eq.(\ref{decoupling}) illustrates that any perturbation away from ideal adiabatic limit tends to affect $T_{CMB,0}$ and hence the age of the Universe inferred from the CMB (and equivalently, $H_0$ estimation).

The age of the Universe has been measured completely independently using the age of the oldest stars in globular clusters of Milky Way showing \cite{2007.06594}
\begin{equation}
    \label{ageGC}
    T_{u,GC}=13.5 ^{+0.16}_{-0.14}(stat.)\pm 0.5 (sys.) \ \mbox{Gyr}.
\end{equation}

This estimate is independent of any cosmological model. 
The two independent measurements Eqs.\eqref{ageT} and \eqref{ageGC} are remarkably consistent, testimony to the essentially adiabatic evolution of the CMB since decoupling. 

Here, we exploit this age consistency (Eqs.\eqref{ageH}-\eqref{ageGC}) to set a novel constraint limiting the interactions of unparticle cosmology with the CMB.

\section{Heat transfer over a Hubble time \label{sec2n}}

One of the key concerns regarding the proposed cosmological models that include additional particles or extra degrees of freedom is the possibility of heat exchange with the CMB. 
Generally, the amount of transferred energy can be calculated by addressing the optical depth of the medium containing the new particles. 
Optical depth describes the transparency level of the medium for photon radiation, conventionally defined as \cite{rybi79}
\begin{eqnarray}
    \label{heat1}
   \tau=\int ^{s}_{s_0}\sigma n\ ds,
\end{eqnarray}
where $\sigma$ is the cross section of the new particles interaction with the photons, $n$ is the number density of the new particles, and $ds$ shows the traveling path of the light rays varying from $s_0$ to $s$.
Optical depth, $\tau$, determines the change in intensity of the light rays due to interactions with the new particles. $\tau$ is also representative for the probability of a single photon having interactions with new particles, if divided by the total number of photons in the optically thin limit ($\tau \ll 1$), which is our case of interest. One photon interacting with new particles, hereby experiences a change in energy 
\begin{equation}
    \label{od1}
    |\Delta E|\simeq E_0 \tau.
\end{equation}

Equivalently in Eq.\eqref{od1}, $\tau$ represents the fractional change in total heat of the CMB. 

Calculating the optical depth of the medium over a Hubble scale, Eqs. \eqref{heat1} gives
\begin{eqnarray}
    \label{od3}
    \tau \cong \sigma n R_H,
\end{eqnarray}
where $R_H \simeq c/H_0 \approx 1.4 \times 10^{26} \mbox{m}$.

In the following we assess Eqs. (\ref{od1}-\ref{od3}) in heat transfer between the CMB and unparticles in light of their scale invariant properties, subject to Eqs.(\ref{ageT}-\ref{ageGC}). 
{In doing so, $\sigma$ refers to an effective cross-section, representative in the mean across the broad mass-spectrum of unparticles, in addition to its energy spectrum defined to leading order by its temperature.}

\section{Unparticle cosmology properties \label{Sec3n}}

Unparticle cosmology is defined by a finite-temperature trace anomaly of the energy-momentum tensor of unparticle stuff, breaking conformal invariance. 
This is parametrized by a $\beta$-function, representing a departure from an IR fixed point \cite{0809.0977}. Alongside the first law of thermodynamics, this gives the following expressions for unparticles energy density and pressure \cite{0809.0977,artimovski}
\begin{eqnarray}
\label{1}
\begin{array}{lll}
\mathcal{\rho}_{\mathcal{U}}=\sigma T^4+A\left(1+\frac{3}{\delta}\right)T^{4+\delta}=\sigma T^4+B T^{4+\delta},\\ \\
\mathcal{P}_{\mathcal{U}}=\frac{1}{3}\sigma T^4+\left(\frac{A}{\delta}\right)T^{4+\delta}=\frac{1}{3}\sigma T^4+\frac{B}{\delta+3}T^{4+\delta}.
\end{array}
\end{eqnarray}
Here, the first term of these two equations is radiation-like with radiation constant $\sigma$ defined in \cite{0809.0977}.  While not quantified in detail, $\sigma \gg a$ is expected, where $a=4\sigma_B/c$ denotes the radiation constant in electromagnetism (EM) in terms of the Stefan-Boltzmann constant $\sigma_B$ and the velocity of light $c$.
 
Eq.(\ref{1}) gives rise to a Friedmann-Robertson-Walker (FRW) universe with unparticles in addition to the standard energy components of, notably, baryonic matter and EM radiation. 
The fact that the radiation constant $\sigma$ of unparticles  is ill-constraint implies that 
\begin{equation}
\label{Bparam}
 B=A(1+\frac{3}{\delta})
\end{equation}
is a free parameter, where $A$ is defined based on $\beta$-function and the coupling of the theory and $\delta=2(d_{\mathcal{U}}-1)$ with $d_{\mathcal{U}}$ to be the (unknown) scaling dimension of unpaticles \cite{0809.0977,MM}. 
In particular, the sign of $B$ is \textit{not} predicted in this unparticle cosmology model. As such, this theory is incomplete, necessitating careful consideration of suitable observation constraints. 

Shown in Eq.(\ref{1}), unparticle cosmology describes a radiation term accompanied by a term which represents a time-like deviation from radiation. Indeed, the second term is a deviation from the first - the opposite of the familiar radiative corrections in the propagator of SM particles. The first term in Eq\eqref{1} is defined as on-shell on the light cone, $ds^2=0$ while the second is time-like with $ds^2>0$. Due to the ambiguity in the sign of the second term it may be propagating forward (normal) or backward (tachyonic) in time, while the total energy density remains positive. As it stands, the unparticle cosmology model leaves this choice of sign wide open, by the unknown sign of $B$. 

With no constraint on the sign of $B$, we resort to discussing $B>0$ and $B<0$ separately. We note that the limit $B=0$ effectively reduces to a problem of (dark) radiation \cite{0810.512,9911165,0203272}, which has been amply discussed in the literature for its potential to relax $H_0$-tension by its contribution to the Hubble expansion in the early Universe \cite{1607.05617,1801.07260}.

 \subsection{Case I: $B>0$}
 It is inferred from Eq. \eqref{1} that the case with $B>0$ puts no restriction on the magnitude of the radiative and non-radiative terms of unparticles energy condition. Therefore the radiation term can be negligible compared to the non-radiative term at low energy scales (late time universe), leaving the non-radiative term as an effective energy density for non-relativistic unparticles. Consequently the equation of state for unparticles can be safely reduced to
\begin{equation}
\label{eos1}
    \dfrac{\mathcal{P}_{\mathcal{U}}}{\rho_{\mathcal{U}}}\simeq \dfrac{1}{\delta +1}.
\end{equation}

Unparticle cosmology with $B>0$ and $-3<\delta<0$ has recently been confronted with late-time $H(z)$-data in the shadow of $H_0$-tension \cite{MM}. While the observational results do not favor this case, $B>0$ is nevertheless found to be a potentially meaningful detour to holographic dark energy, 
\begin{equation}
\label{MM}
\delta=-2.06 \pm 0.46,
\end{equation}
being the best fit value in confrontation with data \cite{MM}. 
Since holography is a non-local theory, Eq. (\ref{MM}) is at the edge of unparticle cosmology. 

 \subsection{Case II: $B<0$}
 Independently, the case with $B<0$ has recently been considered in \cite{2010.02998}, {\em assuming} interactions between unparticles and SM particles can be completely ignored. Focusing on the range $-3<\delta<0$, most relevant to unparticle cosmology.\\
\indent In this case the first term in Eq. \eqref{1} is responsible for positive energy density meaning that the radiation component of unparticles is always dominant over the non-radiative term. Therefore, the equations of state should be considered in its complete form. This, in its own right, prevents unparticles to have pure non-relativistic characteristics.\\
\indent Nevertheless, according to Eq. (\ref{1}) $B<0$ unparticle cosmology is protected against tachynoc behavior overall by a threshold temperature $T_c$ for which $\rho_{\mathcal{U}}=-P_{\mathcal{U}}$, satisfying \cite{artimovski,2010.02998}
\begin{eqnarray}
\label{tc}
T_c=\left[\frac{4}{3} \left(\frac{-\sigma}{B}\right)\left(\frac{\delta+3}{\delta+4}\right)\right]^{1/\delta}. 
\end{eqnarray}
The case with $B>0$ and $-4<\delta<-3$ would also be capable of introducing a $T_c$, but this is ruled out by violating the unitarity of the model \cite{MM}.

\begin{center}
\begin{figure}[!htb]
\includegraphics[width=0.9\linewidth]{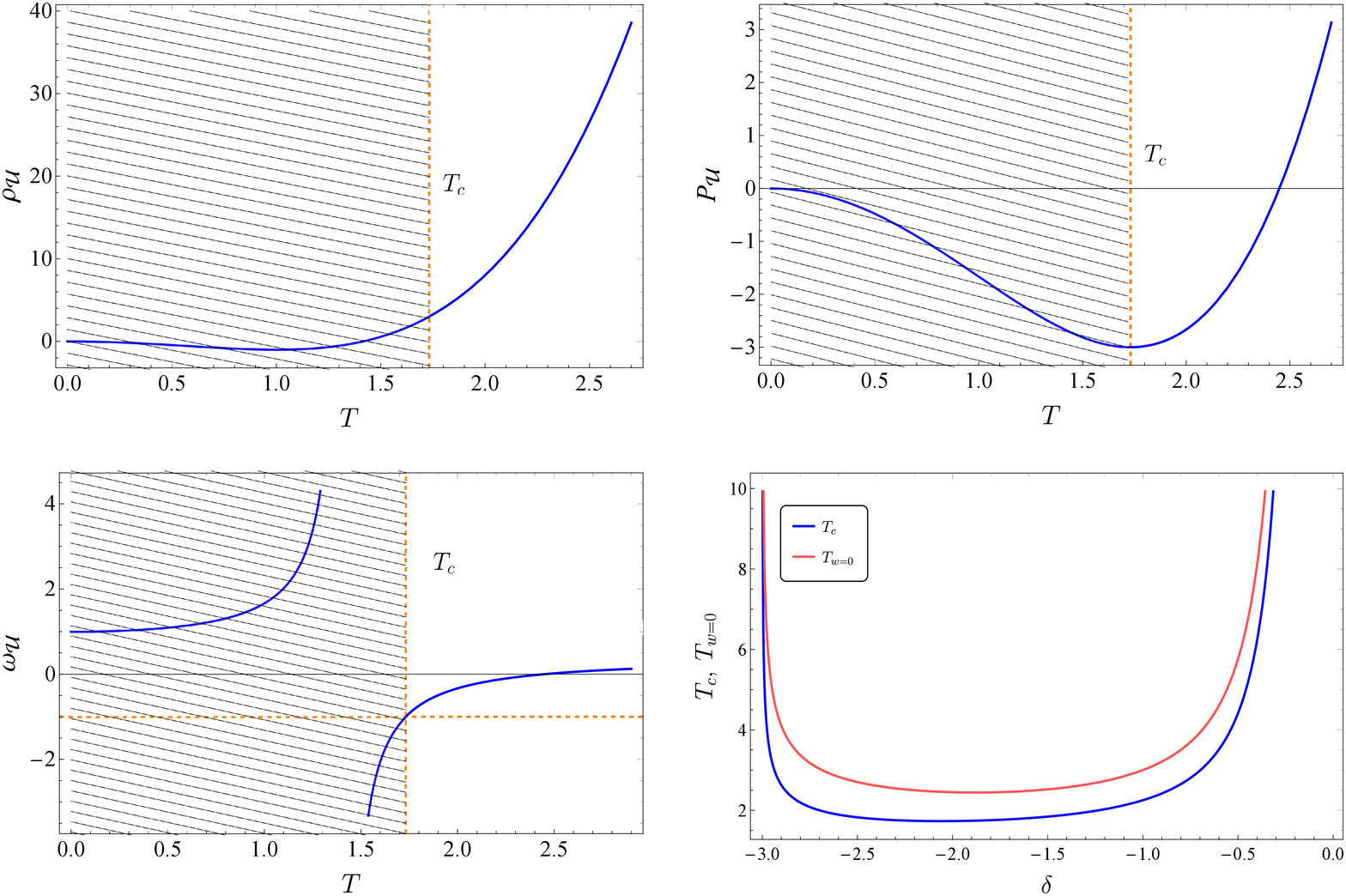}
\caption{{\scriptsize Schematic evolution of $\rho_{\mathcal{U}}$ (top left panel), $P_{\mathcal{U}}$ (top right panel) and $\omega_{\mathcal{U}}=P_{\mathcal{U}}/\rho_{\mathcal{U}}$ (bottom left) assuming $B=-2$ and, motivated by \cite{MM}, $\delta=-2$ in Eq. \eqref{MM}. The vertical orange line shows $T_c$ where $P_{\mathcal{U}}=-\rho_{\mathcal{U}}$. $T_c$ secures the model from having tachyonic behavior i.e. negative energy density for unparticles. The bottom right panel shows the evolution of $T_c$ and $T_{w=0}$ versus $\delta$. We note that the ratio $T_{w=0}/T_c \in (1.2,1.6)$ is slowly decreasing over $-3<\delta<0$. }}\label{fig:figure 1}
\end{figure} 
\end{center}

In \cite{2010.02998}, $T_c$ is identified with a floor in the temperature, effectively introducing a cosmological constant, when pressure has turned negative (Fig. \ref{fig:figure 1}). Total unparticle energy density $\rho_{\mathcal{U}}$ in Eq. \eqref{1} remains positive for

\begin{equation}
    \label{Ti}
    T\ge T_c>T_0,
\end{equation}
protecting it against a zero-crossing at $T_0=\left( \frac{-\sigma}{B} \right)^{\frac{1}{\delta}}$.
 
We note that the temperature at which unparticles behave as dark matter, satisfying $P_\mathcal{U} =0$ in Eq. (\ref{1}) with $w=0$, is
\begin{equation}
 \label{tw}   
 T_{w=0} = \left (\dfrac{4}{4+\delta} \right )^{-1/\delta} T_c,
\end{equation}
which for $-3 < \delta < 0$ gives $1.2\times T_c \lesssim T_{w=0} \lesssim 1.6\times T_c$, illustrating $T_{w=0}$ to be above $T_c$. Since Eq. (\ref{tw}) is independent of \textit{B}, therefore, 
$ T_0 < T_c < T_{w=0}$ holds for both $B > 0$ and $B < 0$. 
The evolution of $T_c$ and $T_{w=0}$ is depicted in the bottom right panel of Fig. \ref{fig:figure 1}, illustrating $T_{w=0}/T_c>1$ and slowly decreasing with $-3<\delta<0$. 

Summarizing the above, we have the temperature inequalities
\begin{equation}
    \label{Tii}
    T_{w=0} > T_c > T_0
\end{equation}
for the cases of unparticle Dark Matter, unparticle Dark Energy and the absolute minimum temperature of unparticles for $B < 0$.

\section{Late-time unparticle cosmology and the CMB \label{Sec4n}}

Unparticle cosmology posits an unconventional cosmological content in unparticle stuff. While radiation-dominated early on, as the universe gets colder, the unparticle admixture Eq. (\ref{1}) effectively describes a tilt of energy from radiation (null) toward the second (time-like) non-radiative term. In case of $B<0$, $T_c>0$ renders this energy transfer finite in asymptotic evolution to the future at which $w=-1$: $P_{\mathcal{U}}=-\rho _{\mathcal{U}}$ {\em assuming ideal} decoupling from SM particles \cite{2010.02998}. In this limit,
the contribution of unparticles reduces to a finite-temperature cosmological constant (Fig. \ref{fig:figure 1}). 

At $T_c$, $w=-1$ gives
\begin{eqnarray}
\rho_{r,\mathcal{U}}=-\frac{3}{4}\left(\dfrac{\delta+4}{\delta+3}\right)\rho_{nr,\mathcal{U}},
\label{EQN_E}
\end{eqnarray} 
where we denote $\rho_{r,\mathcal{U}}=\sigma T^4$ and $\rho_{nr,\mathcal{U}}=BT^{4+\delta}$ for the radiation and, respectively, non-radiation term of the energy density of unparticles Eq. (\ref{EQN_E}). By Eq. (\ref{EQN_E}), $w=-1$ describes an approximate equipartition between $\rho_r$ and $\rho_{nr}$ in $-3<\delta<0$. \\
\indent Eq. \eqref{EQN_E} guarantees that $\rho _{r, \mathcal{U}}$ is always greater than $\rho _{nr,\mathcal{U}}$, making it reliable to estimate the order of magnitude of $\rho _{\mathcal{U}}$ to be approximately equal to $\rho _{r,\mathcal{U}}$.

If at high energies, the radiation constant $\sigma$ in Eq. \eqref{1} is similar to the high-energy BZ limit \cite{0809.0977}, then
\begin{eqnarray}
\sigma={\cal O}\left(10^2a\right)
\label{EQN_s}
\end{eqnarray}
associated with a number of degrees of freedom far in excess of the two polarizations of EM in our standard model.
This possibly large number of degrees of freedom is anticipated by the high-temperature limit 
\begin{equation}
    \label{BZ}
\rho _{BZ}= \dfrac{3}{\pi ^2}g_{BZ} T^4,
\end{equation}
where $g_{BZ}$ represents the relativistic degrees of freedom of BZ fields. According to \cite{0809.0977}, $g_{BZ}=100$ and $g_{\mathcal{U}} = g_{IR} \geq \mathcal{O}(100)$. 

Neglecting interactions between unparticles and SM particles, Eq. (\ref{EQN_E}) permits an estimate of $T_c$ that may be compared with the temperature $T_{CMB}\simeq2.73$K of the CMB with Planck density $\Omega _{CMB,0}=8.5 \times 10^{-5}$, satisfying
\begin{equation}
    \label{EM}
\rho_{EM}=\left(\frac{\pi ^2}{15}\right) T^4\simeq 0.658  T^4,    
\end{equation}
in natural units ($\hbar=c=1$). With $\Omega _{\mathcal{U}}\simeq 1$ in the state $w=-1$,
\begin{equation}
\eta=\dfrac{\Omega_{\mathcal{U}}}{\Omega_{CMB}}\simeq 10^4. 
\label{EQN_eta}
\end{equation}

Combining Eq. (\ref{BZ}-\ref{EQN_eta}) and assuming $g_{\mathcal{U}}\simeq 100$ gives $T_c$ in comparison with CMB temperature
\begin{equation}
    \label{Tc}
    T_c\simeq \left( \frac{10^4 \pi ^4}{45 g_{\mathcal{U}}}\right)^{1/4} T_{CMB}\simeq 4\ T_{CMB}.
\end{equation}

Thus, unparticles are considerably warmer than the CMB ($1\, \mbox{TeV} \gg T_c>T_{CMB}$). It is notable that their temperature ($T_c$) is greater than $T_{CMB}$ by a factor of a few, which is secure because of the power of $\frac{1}{4}$ in Eq. (\ref{Tc}).

Including much different energy densities of Unparticles and CMB photons (Eq. (\ref{EQN_eta})): \textit{unparticles pose a potentially enormous heat exchange with the CMB upon retaining finite interactions with the CMB photons.}

However, according to Sec.\ref{sec1n}, the $T_{CMB}$ is independently constrained by the current age of the Universe according to the age of the oldest stars in globular clusters to within about 4\% \cite{1902.07081,2007.06594}.  Raising the CMB temperature by warm unparticles in excess of a few percent would lower the age of the Universe below this independent astronomical age estimate \cite{2007.06594}. We certainly need a mechanism to avoid raising the CMB temperature substantially, lest we fall back to Hubble's original estimate of $H_0$ in 1929 \cite{Bahcall,freedman}.\\
\indent For concreteness in preserving the consistency in Eqs. \eqref{ageT} and \eqref{ageGC} within the uncertainty range of Eq. \eqref{ageGC}, we consider a fiducial $4\%$ threshold of heat transfer in Eq. \eqref{od1} between unparticles and the CMB. That is, $\tau \lesssim 0.04$, giving
\begin{eqnarray}
\label{6}
0.04\gtrsim n_{\mathcal{U}} \sigma_{\gamma \mathcal{U}}R_H,
\end{eqnarray}
\noindent where $n_{\mathcal{U}}$ and $\sigma_{\gamma u}$ are unparticles number density and the photon-unparticles cross section, respectively. Eq. \eqref{6} stipulates that unparticles are optically thin over a Hubble scale. The case $\tau >1$ is ruled out as it would imply a thermal equilibrium, effectively raising the CMB temperature by a factor of four according to Eq. \eqref{EQN_eta}, in conflict with Eqs. (\ref{ageT}-\ref{ageGC}).\\ 
\indent Eq. \eqref{6} puts an upper bound on the opacity of unparticles. With known unparticles number density this equation would give us an estimate of unparticles cross section with the CMB photons, but because of the incompleteness of the theory and unparticles undetermined mass by their scale invariance, there are ambiguities in the precise unparticles number density.In the following we assess two marginal cases for $n_{\mathcal{U}}$ and consequently $\sigma _{\gamma \mathcal{U}}$.

\section{Unparticles cross section estimation with the CMB \label{Sec5n}}

Unparticle phase of Banks-Zaks fields can emerge at the energy scales below $1\, \mbox{TeV}$ \cite{0808.0523} giving unparticles the probability of having relativistic or non-relativistic properties according to their mass.

Inherent to unparticles, however, is that their mass, $m_{\mathcal{U}}$, while nonzero, is not known if not undetermined and possibly not unique in the face of multiple species making a family of unparticles. Their relativistic limit is defined by $m_{\mathcal{U}}c^2\ll k_BT$, where $k_B$ is Boltzmann constant in which case the rest mass can be ignored. In contrast, they are non-relativistic if $m_{\mathcal{U}}c^2\gg k_BT$, meaning that rest mass energy is largely dominant over thermal energy. To derive the cross section of unparticles with CMB photons, number density of unparticles should be assessed independently in each regime.

Given the lack of information about unparticles mass, we turn to energy densities to identify if they can be fitted in to these two regimes for the case of $B<0$.

Note that unparticles can not be non-relativistic as it makes the non-radiative term dominant over the radiation term, pushing their energy density to negative values. At the same time, the approximate equipartition in Eq. \eqref{EQN_E} suggests that unparticles might have mildly non-relativistic properties with $m_{\mathcal{U}} c^2 \sim k_BT$.

In what follows we recall that the CMB photons number density is $n_{\gamma}=420\ cm^{-3}=4.2 \times 10^8 m^{-3}$ with 
energy density $\rho_\gamma \simeq 0.26\left(1+z\right)^4$\,eV\,cm$^{-3}$ \citep[e.g.][]{2014MNRAS.438.2065C}.

\subsection{Relativistic limit\label{Sec5n-1}}
In the relativistic limit $k_BT\gg m_{\mathcal{U}}c^2$ and number density of any relativistic species in thermal equilibrium can be estimated by
\begin{equation}
    \label{Rn}
    n_{\mathcal{U}}^{eq}=\dfrac{\zeta (3)}{\pi ^2}g_{\mathcal{U}}T^3,
\end{equation}
where $g_A$ is the number of relativistic degrees of freedom. \\
\indent Combining Eqs. \eqref{Tc} and \eqref{Rn} for unparticles and CMB photons gives\\
\begin{equation}
    \label{RnU}
    \dfrac{n_{\gamma}}{n_{\mathcal{U}}}=\dfrac{g_{\gamma} T^3_{CMB}}{g_{\mathcal{U}}T^3_{c}}.
\end{equation}
Where $g_{\gamma}=2$ and the aggregate relativistic degrees of freedom of the whole unparticles family $g_{\mathcal{U}}=100$, giving $n_{\mathcal{U}}\simeq 1.3 \times 10^{12} \ \mbox{m}^{-3}$ as an approximate scale of unparticles number density at relativistic regime. Combining Eq. (\ref{6}) and Eq. (\ref{RnU}), we conclude
 \begin{equation}
 \label{7-1}
 \sigma _{\gamma \mathcal{U}} \lesssim 10^{-40}\ \mbox{m}^2=10^{-3}\ \mbox{nb}. 
 \end{equation}

To put Eq. \eqref{7-1} into the context we recall that $\sigma _{\gamma \gamma} \simeq 10^{-35}\ \mbox{m}^2=100\ \mbox{nb}$ \cite{1904.01243,1810.04602,1305.7142,9709415} and $\sigma_{\gamma \nu} \sim 10^{-43}\ \mbox{m}^2=10^{-6} \ \mbox{nb}$ at high energy levels \cite{Dicus93,0012257,0511042}. Although we have no exact measurement of neutrino-photon interactions, nevertheless Eq. (\ref{7-1}) puts unparticles possibly at the edge, now, of our standard model of particle physics.\\

\subsection{Mildly non-relativistic limit\label{Sec5n-2}}

Unparticles can be assessed as mildely non-relativistic species either due to their possibly large masses or low energy scale regime, satisfying $k_B T \sim m_{\mathcal{U}}c^2$. Pushing the limit down to but not lower than what is permitted by equipartition (in the sense that full equipartition in Eq. \eqref{EQN_E}, $\rho_{\mathcal{U}}=0$) and with reference to Eqs. \eqref{Tc}-\eqref{Ti}, the kinetic energy and total energy of a non-relativistic scalar unparticle with one degree of freedom are $K_{\mathcal{U}}=(1/2)k_BT_{\mathcal{U}}$ and $E_{\mathcal{U}} \sim k_{B}T_{\mathcal{U}}$, respectively.

Abandoning Eq.\eqref{Rn}, we now appeal to number density in the mean
\begin{equation}
    \label{NRn}
    n_{\mathcal{U}}=\dfrac{\rho _{\mathcal{U}}}{E_{\mathcal{U}}}\sim \dfrac{\rho_{\mathcal{U}}}{k_BT_{c}},
\end{equation}
giving an estimate of non-relativistic number density of unparticles. 

Compared with CMB photons number density, once more, we find
\begin{equation}
\label{nu}
n_{\mathcal{U}}=\dfrac{\rho_{\mathcal{U}}}{E_{\mathcal{U}}} \simeq 10^4\ \dfrac{\rho_{CMB}}{k_BT_{c}} \simeq 2.5 \times 10^3\ n_{\gamma},
\end{equation}
where the last equality is based on Eq. \eqref{Tc}, i.e. $n_{\mathcal{U}} \simeq  10^{12} \ \mbox{m}^{-3}$. Combining the bound in Eq. (\ref{6}) and Eq. (\ref{nu}), we conclude
 \begin{equation}
 \label{7-2}
 \sigma _{\gamma \mathcal{U}} \lesssim 10^{-40}\ \mbox{m}^2=10^{-3}\ \mbox{nb}. 
 \end{equation}
 
\section{Conclusions\label{Sec6n}}

Unparticle cosmology suggests some novel avenues to address accelerated Hubble expansion in the late-time Universe. 
The indeterminate sign of $B$ in Eq.(\ref{1}), however, prompts us to pursue a confrontation with observations for $B>0$ and $B<0$ separately. 
While the case $B>0$ - normal time-like deviation from radiation - can be confidently ruled out \cite{MM}, the case $B<0$ poses a different cosmological scenario characterized by a floor in the temperature $T_c$, e.g. \cite{2010.02998}. $T_c$ protects unparticle cosmology against tachyonic behavior and introduces a late-time evolution very similar to $\Lambda\mbox{CDM}$ (Fig. \ref{fig:figure 1}). 

In fact, as indicated by the temperature inequalities Eq.\eqref{Tii}, unparticle cosmology has a potentially complex phenomenology taking us through DM-like and DE-like phases as the temperature decreases. Regardless a basic premise retains that  a non-vanishing but possibly small interaction with SM particles remains down to arbitrarily low energies.

Specifically for DE-like unparticles, we find $1\, \mbox{TeV} \gg T_c\simeq 4T_{CMB}$ (Eq. (\ref{Tc})). Combined with $\Omega_{\cal U}\simeq 10^4\ \Omega_{CMB}$ (Eq. (\ref{EQN_eta})), the model prediction is that the CMB is exposed to an enormous heat bath. This exposure is suppressed only by a sufficiently small cross section of unparticles (Eqs.(\ref{7-1}) and \eqref{7-2}).

The theory of unparticles is incomplete in the sense that it does not determine for instance the sign of $B$ and a range of unparticles mass. 
Unparticles might be considered a family of species with different masses that collectively contribute to the evolution of the universe. 
Accordingly, unparticles may be a mixture of relativistic and mildly non-relativistic species. 
The consistency between Eqs. \eqref{7-1} and \eqref{7-2} suggests our cross section bounds to be relatively robust {with minimal assumptions on their microphysical origin.}

Preserving weak interactions with SM particles \cite{0703260}, we find unparticles to be a potential candidate for warm Dark Matter and/or interacting Dark energy (Eq. \eqref{Tii}). Their minuscule cross section in Eqs. (\ref{7-1}) and \eqref{7-2} yet again put unparticle cosmology to the edge, now of our standard model interactions with the CMB photons. Though small, these may nevertheless have phenomenological implications. For instance, it is shown that a tiny non-gravitational interaction between DM and SM particles can alter the speed of structure formation and therefore be relevant to Missing satellites problem \cite{1404.7012,1412.4905}.

{In the approximation of an effective cross section $\sigma_{\gamma{\cal U}}$, we apply the conventional expression for opacity Eq.\eqref{heat1} 
to impose a bound on the relative temperature shift in the CMB allowed by the uncertainty of a few percent in the astronomical age of the Universe.}

{Considerably sharper constraints cross-sections may derive from spectral distortions \citep{Zeld1969}, given the COBE-~FIRAS bound $y\le 1.5\times 10^{-5}$ 
\citep{Fix1996} on the Compton $y$-parameter with associated temperature shift \citep[e.g.][]{Fix1996,Chluba2014}
\begin{eqnarray}
\frac{\Delta T}{T}\simeq 4y,
\label{EQN_y}
\end{eqnarray}
for a relative energy transfer $\Delta U/U \simeq \Delta T/T$ in photon-number preserving weak Compton up-scattering \citep[e.g.][]{Ali2015,Zott2016} of the CMB, here applied to unparticles in late-time cosmology, Eq.\,(\ref{EQN_y}). 
Specifically, the correlation Eq. \eqref{EQN_y} assumes a conventional constant cross-section across the continuous mass and energy spectrum of unparticles, i.e., a vanishing gradient with respect 
to mass and energy. (For a discussion on a power-law energy-dependent cross-section, see, \cite[e.g.][]{Ali2015}.)  
If applicable, the resulting bound on $\sigma_{\gamma{\cal U}}$ is then on par with $\sigma_{\gamma\nu}$ (recalled in \S6.1), 
i.e., at the very edge of the standard model.\\
\indent Note, however, that an assumed constant cross-section may not hold for unparticles. After all an intrinsic scale $\sigma_0$ of their cross-section may not exist if conformal invariance of unparticles is universal beyond their mass alone. In this case, it is emergent and dimensional analysis suggests, for instance,

\begin{equation}
    \label{dcs} 
    \sigma=\frac{A}{E\times m},
\end{equation}
\noindent over distributions of the unparticles energy $E$ and mass $m$, where $A$ is a small dimensionless constant (working in natural units). Here, evidently $E<\Lambda_{\mathcal{U}}$, where $\Lambda_{\mathcal{U}}$ is the energy scale of the phase transition discussed in \S \ref{sec:intro}. This energy cut-off is called for, since $\sigma \rightarrow 0$ in Eq. \eqref{dcs} as $E \rightarrow \infty$ is at odds with the standard model interactions recovered at $E>\mathcal{M}_{\mathcal{U}}>\Lambda_{\mathcal{U}}$. \\
\indent Alternatively, though perhaps less likely, unparticles do carry an intrinsic scale $\sigma_0$ about some fiducial values of mass and respectively energy $m_0$ and $E_0$. In this case, a more general scenario might apply like

\begin{equation}
    \label{MCS} 
    \sigma=\sigma_0 \left( \frac{m}{m_0}\right)^{p}\left( \frac{E}{E_0}\right)^{q}
\end{equation}

\noindent  with (unknown) power indices $p$ and $q$. If so, Eq. \eqref{MCS} is a generalization of Eq. \eqref{dcs}, reducing to it when $p=q=-1$. Further exploration of Eqs. \eqref{dcs}-\eqref{MCS} is left for a future investigation.}


\section*{Acknowledgment}
The authors would like to thank the anonymous reviewer for critical and constructive comments on this work. 
This research is supported by NRF of Korea Nos. 2015R1D1A1A01059793, 2016R1A5A1013277, and 2018044640.

 \bibliographystyle{elsarticle-num} 
 \bibliography{mybibfile}





\end{document}